\documentclass[journal,twoside,web]{ieeecolor}
\usepackage{generic}
\usepackage{cite}
\usepackage{amsmath,amssymb,amsfonts}
\usepackage{algorithmic}
\usepackage{graphicx}
\usepackage{textcomp}
\usepackage{textgreek}
\usepackage{xcolor}

\def\BibTeX{{\rm B\kern-.05em{\sc i\kern-.025em b}\kern-.08em
    T\kern-.1667em\lower.7ex\hbox{E}\kern-.125emX}}
\begin{document}
\title{Temperature Dependent Current Dispersion Study in $\beta$-Ga$_2$O$_3$ FETs Using Sub-Microsecond Pulsed IV Characteristics}
\author{Abhishek Vaidya, and Uttam Singisetti
\thanks{We acknowledge the support from Air Force Office of Scientific Research under award number FA9550-18-1-0479 (Program Manager: Ali Sayir) and from NSF under awards  ECCS 1607833 and ECCS - 1809077. This work was performed partly at Shared Instrumentation Lab, University at Buffalo. This research used resources of the Center for Functional Nanomaterials, which is a U.S. DOE Office of Science Facility, at Brookhaven National Laboratory under Contract No. DE-SC0012704 \& Center for Functional Nanomaterial, Brookhaven National Lab.}
\thanks{Abhishek Vaidya \& Uttam Singisetti are with Electrical Engineering Department, University at Buffalo, Buffalo, NY - 14260}

}

\maketitle

\begin{abstract}

A comprehensive study of drain current dispersion effects in $\beta$-Ga$_2$O$_3$ FETs has been done using DC, pulsed and RF measurements.
Both virtual gate effect in the gate-drain access region and interface traps under the gate are most plausible explanations for the experimentally observed pulsed current dispersion and high temperature threshold voltage shift respectively.
Unpassivated devices show significant current dispersion between DC and pulsed IV response due to gate lag effect characterized by time constants in the range of 400~$\mu s$ to 600~$\mu s$. 
An activation energy of 99~$meV$ is estimated from temperature dependent Arrhenius plots. A variable range hopping based slow transport in conjunction with the observed shallow trap level is attributed to the observed slow transient response of drain current with respect to time.
Reactive ion etching step during the device fabrication is most likely responsible for introducing the traps. 
Effect of traps can be minimized by using surface passivation layers, in this case, Silicon Nitride which shows significant improvement in the current dispersion and RF cutoff frequency.
This work demonstrates the detrimental effect the traps can have on the current dispersion which significantly limits the high frequency operation of the device.

\end{abstract}

\begin{IEEEkeywords}
$\beta$-Ga$_2$O$_3$, current dispersion, Interface traps, Pulsed IV, Virtual gate
\end{IEEEkeywords}

\section{Introduction}
\label{sec:introduction}
\IEEEPARstart{B}{eta} Gallium oxide has garnered a lot of interest of researchers across the globe because of it's attractive material properties such as ultra-wide bandgap of \texttildelow4.8-4.9~$eV$ \cite{Tippins65,Varley10}, good electron mobility, high breakdown field strength of 8~$MV/cm$ \cite{Konishi17,Higashiwaki12,Kohei12,Chabak_APL_16}, and also the ease of growing high quality epitaxial films with controllable doping \cite{joishi2018low,bhuiyan2020mocvd,anhar2019mocvd,sasaki2012device,murakami2014homoepitaxial,goto2018halide,zhang2019mocvd,han2018n} on melt grown bulk crystals. The large breakdown field strength translates to high Baliga's Figure of Merit (\textit{BFoM}) making it superior candidate for power device applications. Several groups have demonstrated kilovolt class field effect transistors (FETs) and Schottky diodes \cite{zeng20181,zeng2019field,hu2018lateral,hu2018breakdown,hu2018enhancement,li20181230,tetzner2019lateral,sharma2020field,hu2018field,mun20192,konishi20171,gong20201,huang20203}. In addition, $\beta$-Ga$_2$O$_3$ also has a high Johnson's Figure of Merit (\textit{JFoM}) owing to the large calculated electron velocity \cite{Ghosh17}, which has been experimentally verified \cite{zhang2019evaluation}. High \textit{JFoM} implies $\beta$-Ga$_2$O$_3$ is also suitable for high frequency devices such as GHz switches and RF amplifiers. Recently a few groups have successfully demonstrated $\beta$-Ga$_2$O$_3$ field effect transistors with current gain cutoff frequencies ($f_t$) in the GHz range \cite{Green17,Chabak_MTT_18, kamimura2020rf}. Xia \textit{et.al.} reported the highest $f_t$ of 27~$GHz$ in a delta doped $\beta$-Ga$_2$O$_3$ (Ga$_2$O$_3$)FET with regrown ohmic contacts \cite{Xia19}.\par

In addition to the cutoff frequencies, another important factor for high frequency performance of the device is DC to RF drain current dispersion which causes transconductance to collapse as the frequency of operation increases with an associated knee-walk off phenomenon. This is primarily caused by interface and/or bulk traps which act as generation and recombination centers. The time constant associated with the traps dictates the amount of dispersion at an given frequency. Transient response analysis using pulsed current-voltage measurements is a popular method to identify the type and location, both physical and energy, of the trap. This method has been widely used to study dispersion in GaN based devices \cite{Faqir08,Meneghesso13,Tirado07,Meneghesso04,Vetury01,Verzellesi}. While, only a few reports exist on Ga$_2$O$_3$ devices. \par

There are two kinds of $\beta$-Ga$_2$O$_3$ RF FET devices reported; namely, (a) the source/drain (S/D) regions are either regrown or ion implanted and (b) the S/D n$^+$ layers grown and then removed from the channel region by reactive ion etching (RIE) which are called recessed gate FETs. The recessed gate process requires channel to be exposed to reactive ion etching which has been identified to introduce plasma damage and interface states \cite{Yatabe15,Mamor08,Huang04,Kim12}. This makes it vital to study the effect of traps and find ways to mitigate their detrimental effects on the device performance.
There have been very few studies specifically on $\beta$-Ga$_2$O$_3$ devices. Moser \textit{et. al.}\cite{Moser20} have reported pulsed large signal power performance of $\beta$-Ga$_2$O$_3$ MOSFET. They have analyzed continuous and pulsed output power to provide evidence of presence of traps for the observed dispersion. In contrast, in a previous report, Moser \textit{et. al.} show pulsed current voltage characterization on a 200 ~$nm$ bulk channel FET which shows no appreciable current dispersion \cite{Moser17}. A pulsed large signal RF performance was reported by Singh \textit{et. al}.\cite{Singh18} which shows negligible DC-RF dispersion with microsecond pulses. McGlone \textit{et. al.} explore buffer traps in a $\delta$ doped $\beta$-Ga$_2$O$_3$ structure using Deep Level Transient spectroscopy and double pulsed I-V measurements \cite{McGlone19,joishi2018effect,mcglone2018trapping}. Joishi \textit{et. al.} report double pulsed current voltage measurement using a $5~\mu s$ pulse width on a \emph{Si} $\delta$-doped $\beta$-Ga$_2$O$_3$ FET which shows drastic improvement in current after \textit{in-situ} passivation \cite{joishi2020deep}. Nevertheless, significant knowledge gap exists in the DC-RF dispersion and mitigation strategies for Ga$_2$O$_3$ FET devices. In this work, we report a comprehensive temperature dependent DC-RF dispersion analysis in Ga$_2$O$_3$ MOSFETs using a minimum pulse width of 200~$ns$. We provide the origin of the DC-RF dispersion and the nature and location of traps along with the mechanism of capture and emission processes. We also first report the effectiveness of silicon nitride  passivation in reducing the DC-RF current dispersion in Ga$_2$O$_3$ MOSFETs.

\section{Experimental Details}
\label{sec:experimental}
\begin{figure}[t]
\centerline{\includegraphics[width=\columnwidth]{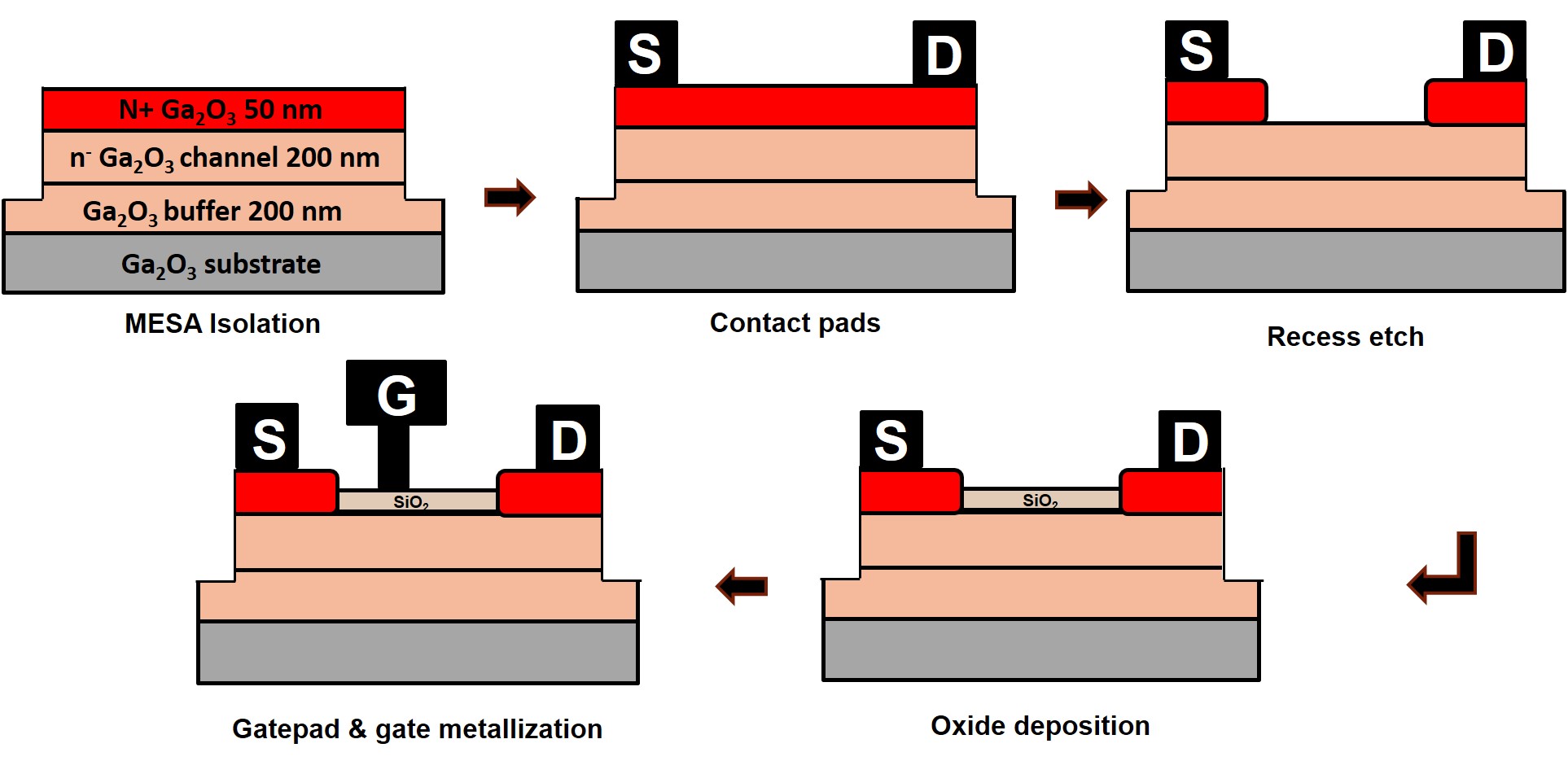}}
\caption{The figure highlights major steps in fabrication process flow of the device in this study.}
\label{fab_flow}
\end{figure}

The semi-insulating Ga$_2$O$_3$ substrates used in this study were grown by edge defined film fed growth method by Tamura Corporation, Japan. Device quality channel and ohmic capping epilayers were grown homoepitaxially on top of 200 ~$nm$ unintentionally doped (UID) Ga$_2$O$_3$ to act as a buffer layer. All the device layers including UID were grown by ozone molecular beam epitaxy (MBE). A gas mixture of ozone and oxygen was used as the oxygen source. The substrate temperature was 700$^{\circ}$C , and the growth rate of Ga$_2$O$_3$ was 0.6 $\mu m/hr$. The channel and ohmic capping layer were grown with target thickness and doping concentrations of $200~nm$~/~$7\times10^{17}~cm^{-3}$ and $50~nm$~/~$1\times10^{19}~cm^{-3}$ respectively. Fig.\ref{fab_flow} shows the schematic of the device fabrication, electron beam lithography (EBL) was used in all the steps. Device isolation etch was first performed using inductively coupled plasma reactive ion etching (ICP RIE) with BCl$_3$/Ar chemistry. Next, Ti/Au/Ni source and drain contacts were deposited by e-beam evaporation followed by a 1 min~/~520$^{\circ}$C N$_2$ anneal to aid the formation of ohmic contacts.
Followed by this, BCl$_3$/Ar gate recess etch was performed between source and drain (L$_{sd}$= 0.5$\mu$m) to remove the highly doped ohmic capping layer. A 20 ~$nm$ plasma ALD SiO$_2$ was deposited as a gate dielectric layer and subsequently annealed at 450$^{\circ}$C to improve its dielectric properties. 
Another lithography was performed to remove the oxide from source and drain contact pads, this time, using CF$_4$/Ar chemistry in ICP RIE. Finally a Ti/Au gate contact pad and gate were deposited to complete the fabrication for the pre-passivation study. SEM micrographs were imaged using Carl Zeiss AURIGA CrossBeam. A 250 ~$nm$ thick plasma enhanced chemical vapor deposition (PECVD) silicon nitride (SiN$_x$) was deposited to study the passivation of the interface states. The SiN$_x$ was removed from the source/drain and gate contact pads to facilitate the probing.\par

\begin{figure}[b]
\centerline{\includegraphics[width =\columnwidth]{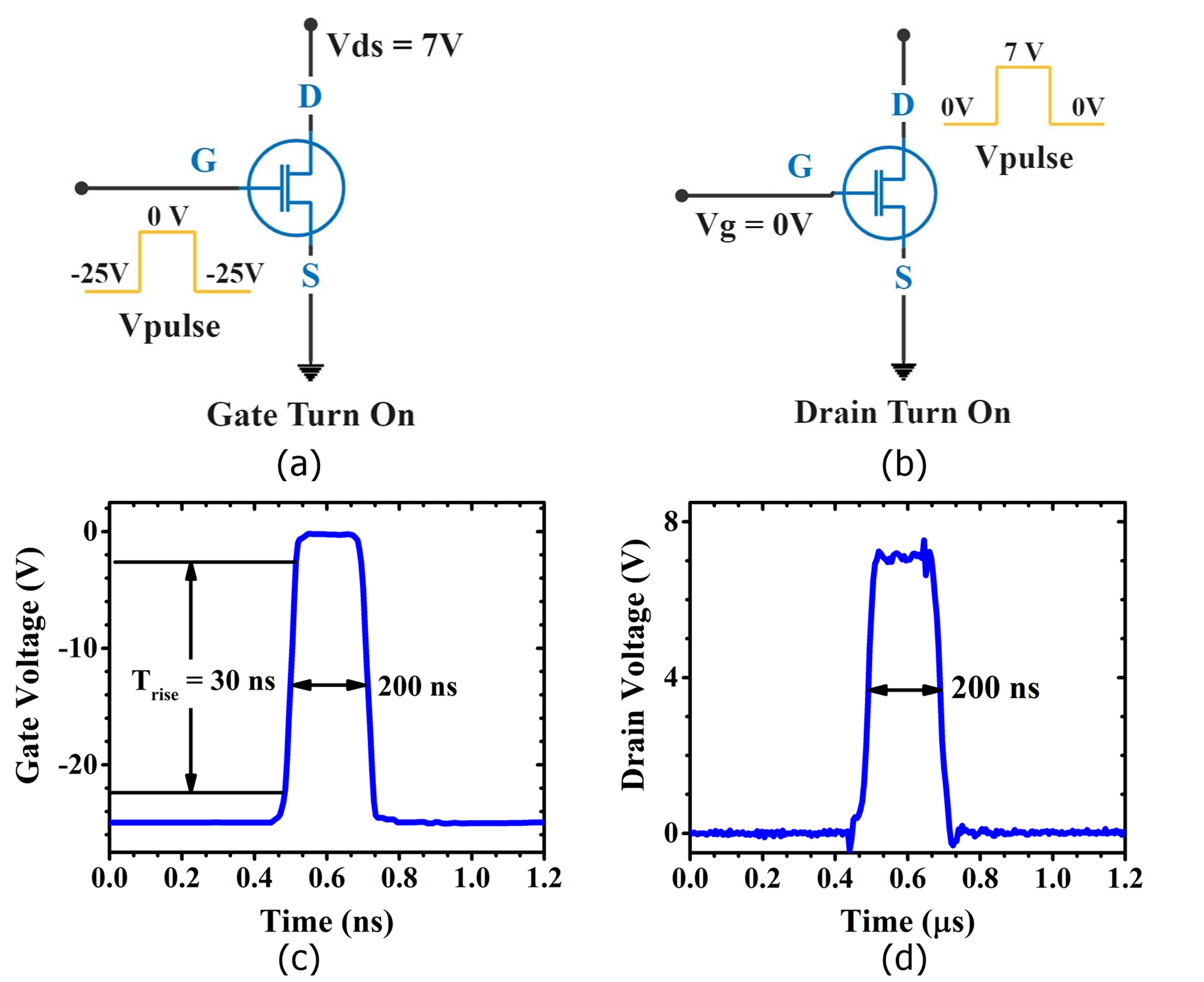}}
\caption{(a) \& (b) show schematic representation of measurement setups for gate and drain turn on measurements respectively. (c) \& (d) show 200~$ns$ gate and drain input voltage pulse shapes.}
\label{pulse_setup_sch}
\end{figure}

For DC current-voltage (I-V) measurements HP4155B semiconductor parameter analyzer was used. A MS-TECH 1000H temperature controlled stage was used to measure up to  300 $^{\circ}$C. Pulsed IV measurements were done using an \textit{Auriga AU-5} high voltage pulsed IV setup capable of sourcing 100~$V$/200~$ns$ at input port and 200~$V$/200~$ns$ at output port with 20~$ns$ rise and fall time. The device was probed using a co-planar waveguide (CPW) RF probes to ensure minimum reflections at the probe-device interface (See Fig. \ref{pulse_setup_sch}).

To analyze the trapping/detrapping response of traps, the well known gate-lag and drain-lag measurement technique was implemented as follows:
\begin{itemize}
    \item \textit{Gate turn-on:} Gate bias is pulsed from off state (V$_g$=-25~$V$) to on state (V$_g$=0~$V$) while maintaining a constant drain bias. Complete IV curves are obtained by sweeping V$_d$ from 0~$V$ to 7~$V$. 
    \item \textit{Drain turn-on:} Drain bias is pulsed from off state (V$_d$=0~$V$) to on state (V$_d$=7~$V$) while maintaining a constant gate bias. Complete IV curves are obtained by sweeping V$_g$ from -25~$V$ to 0~$V$.
\end{itemize}

\begin{figure}[!t]
\centerline{\includegraphics[width=\columnwidth]{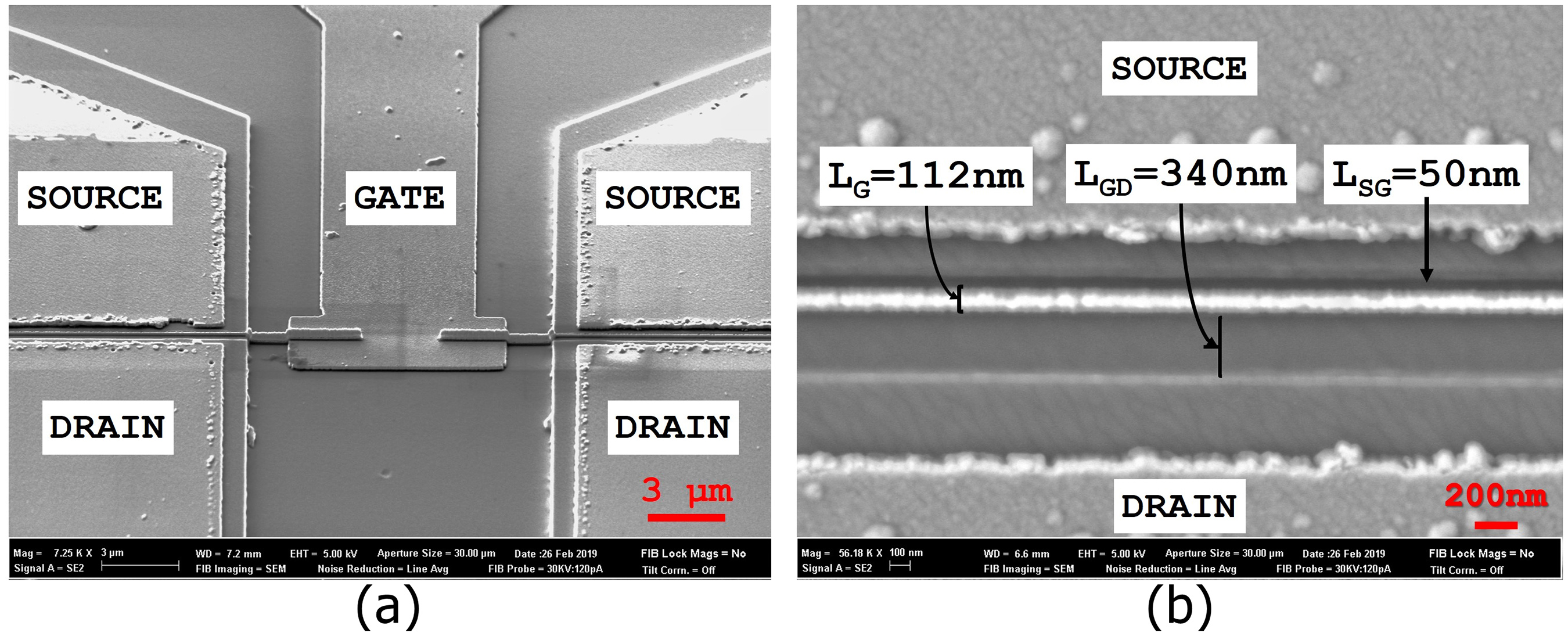}}
\caption{(a)SEM micrograph of the device showing large source and drain contact pads.(b) Zoomed in sem view of the gate recess region showing intrinsic part of the device.}
\label{sem}
\end{figure}
For each study, pulse width was varied from 200~$ns$ to 1~$ms$ with a pulse period of 20~$ms$ to allow sufficient time for the traps to regain their original state after every on pulse. The low duty cycle also ensures device self-heating effects are minimized.
The number of samples were varied between 40 and 2000 based on the pulse width with highest samples for 1~$ms$ pulse and lowest samples for 200~$ns$ pulse.
For each pulsed measurement, the recorded I$_d$ value was averaged over last 10\% time window of the pulse. The measurements were repeated for high temperatures up to 200$^{\circ}$C before and after passivation.\par

S-parameter measurements were carried out using \textit{Agilent E5071} series ENA. The measurements were repeated after passivation.

\section{Results and Discussions}
\label{sec:results}
 The final dimensions of the fabricated device are; gate length (L$_g$) = 112~$nm$, Source-Gate spacing (L$_{sg}$) = 50~$nm$ and Gate-Drain spacing (L$_{gd}$) = 340 ~$nm$. Fig.\ref{sem} shows SEM micrograph of the fabricated device.

\subsection{DC I-V Analysis}\label{subsec:staticIV}

\begin{figure}[!b]
\centerline{\includegraphics[width=\columnwidth]{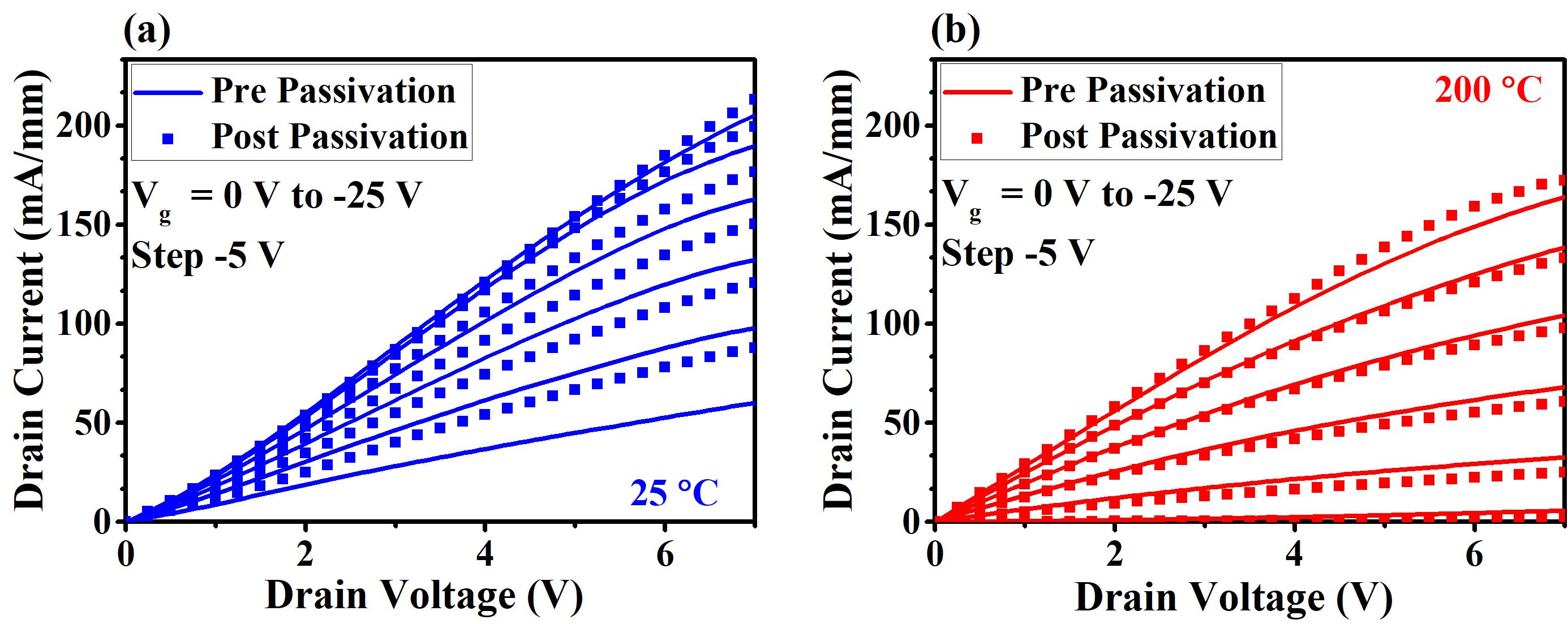}}
\caption{(a) Static I$_d$-V$_d$ comparison before the passivation (line) and after passivation (symbol) at room temperature. (b)Static I$_d$-V$_d$ comparison before the passivation (line) and after passivation (symbol) at 200$^{\circ}$C.}
\label{dc_id_vd_rt_200}
\end{figure}

\begin{figure}[t]
\centerline{\includegraphics[width=\columnwidth]{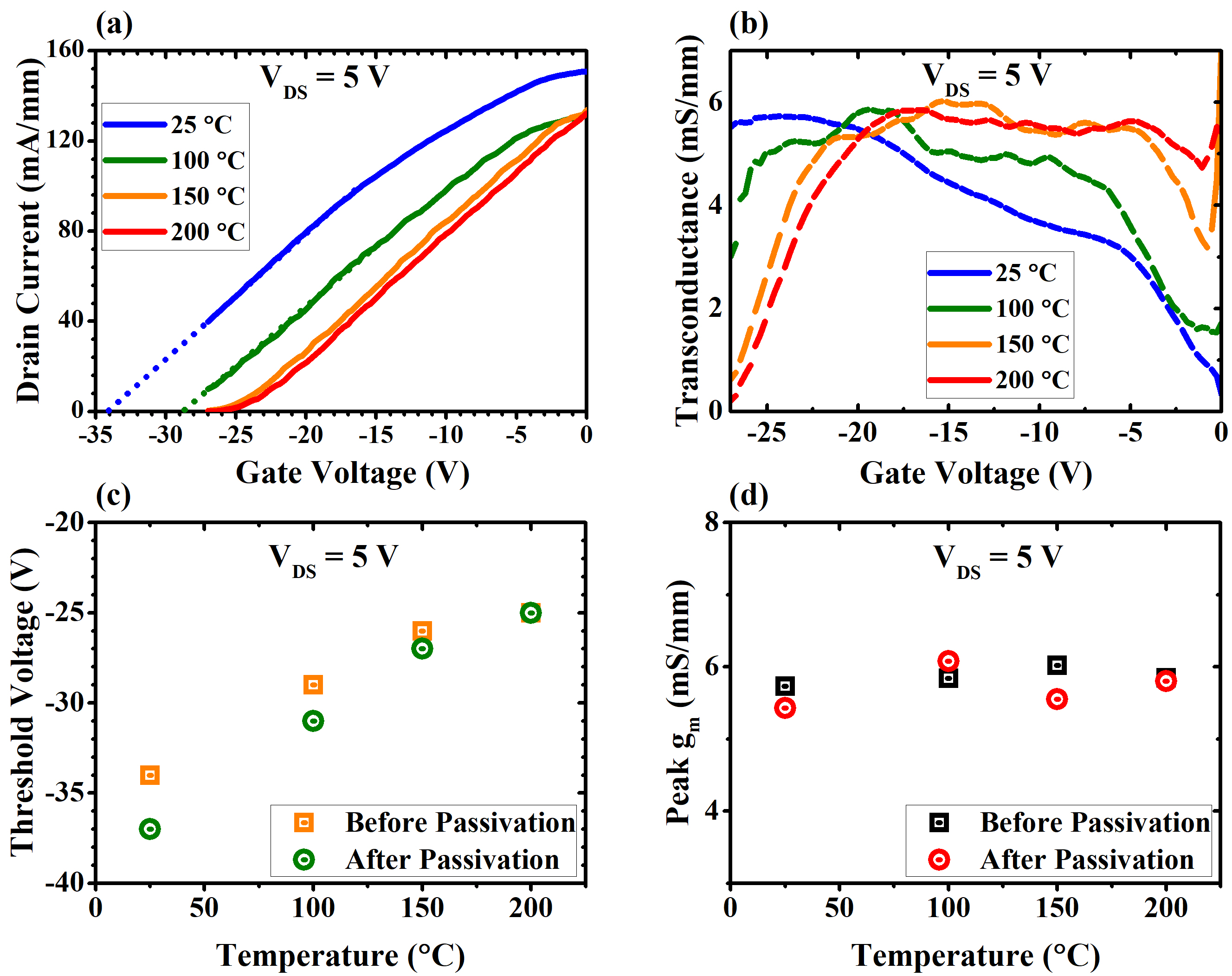}}
\caption{(a) I$_d$-V$_g$ (left) and corresponding (b) g$_m$-V$_g$ (right) plots before passivation at room temperature (blue), 100$^{\circ}$C (green), 150$^{\circ}$C (orange), and 200$^{\circ}$C (red). (c) V$_t$ shift with temperature data before and after passivation shows a similar monotonic positive shift in V$_t$ (d) Peak g$_m$ vs Temperature data before and after passivation indicates little to no change in maximum static g$_m$.}
\label{dc_id_gm_vt}
\end{figure}

Fig. \ref{dc_id_vd_rt_200}(a) and (b) show the measured I$_d$-V$_d$ characteristics at room temperature and 200$^{\circ}$C respective. At room temperature we see incomplete turnoff even at V$_g$=-25$V$. It is attributed to the moderately high doping in the channel, a thicker channel and a smaller gate length resulting in a reduced control over the channel current. Gate voltages lower than -25~$V$ were avoided to protect devices from gate dielectric electrical breakdown. The short channel effects are also evident due to shorter gate length. A maximum current of 210 $mA/mm$ was recorded for this device at V$_g$=0$V$. Non-ideal behaviour of source-drain contacts can also be seen in the graph which is a consequence of non-optimized contact formation process. This also results in higher source and drain resistances which reduces the net drain current and transconductance. The device shows improved ON/OFF ratio at 200$^{\circ}$C for the same V$_g$ as indicated in Fig. \ref{dc_id_vd_rt_200}(b) (solid red line) which is due to a threshold voltage shift that can be attributed to mobile trapped charges in the gate oxide. Contact linearity is improved  at this temperature most likely due to reduced thermionic emission barrier. The figures also show that there is a small change in  DC I-V characteristics after SiN$_x$ passivation.\par
Fig. \ref{dc_id_gm_vt}(a) and (b) show temperature dependent I$_d$-V$_g$ and g$_m$-V$_g$ plots respectively, at V$_{ds}$ = 5~$V$, before passivation. The positive shift in the threshold voltage is evident again suggesting presence of trapped charges in the gate oxide. A maximum g$_m$ of 6~$mS/mm$ has been recorded which as discussed earlier is a considerably lower value for designed device dimensions due to higher source and drain access resistances. As seen in the temperature dependent threshold and peak g$_m$ plots in Fig. \ref{dc_id_gm_vt} (c) \& (d), the passivation does not reduce the threshold voltage shift with temperature. This is again due to the fact the traps under the gate are unaffected by passivation. The peak value of g$_m$ also remains largely unaffected after passivation.

\subsection{Pulsed I-V Analysis}\label{subsec:pulsedIV}

As described in the Introduction, pulsed I-V analysis is important to study the DC to RF dispersion that arise from the dynamic response of traps with change in bias. We carried out drain turn on and gate turn on pulsing individually to isolate the effects of interface traps and bulk traps and localize the source of current dispersion.\par

\begin{figure}[t]
\centerline{\includegraphics[width=\columnwidth]{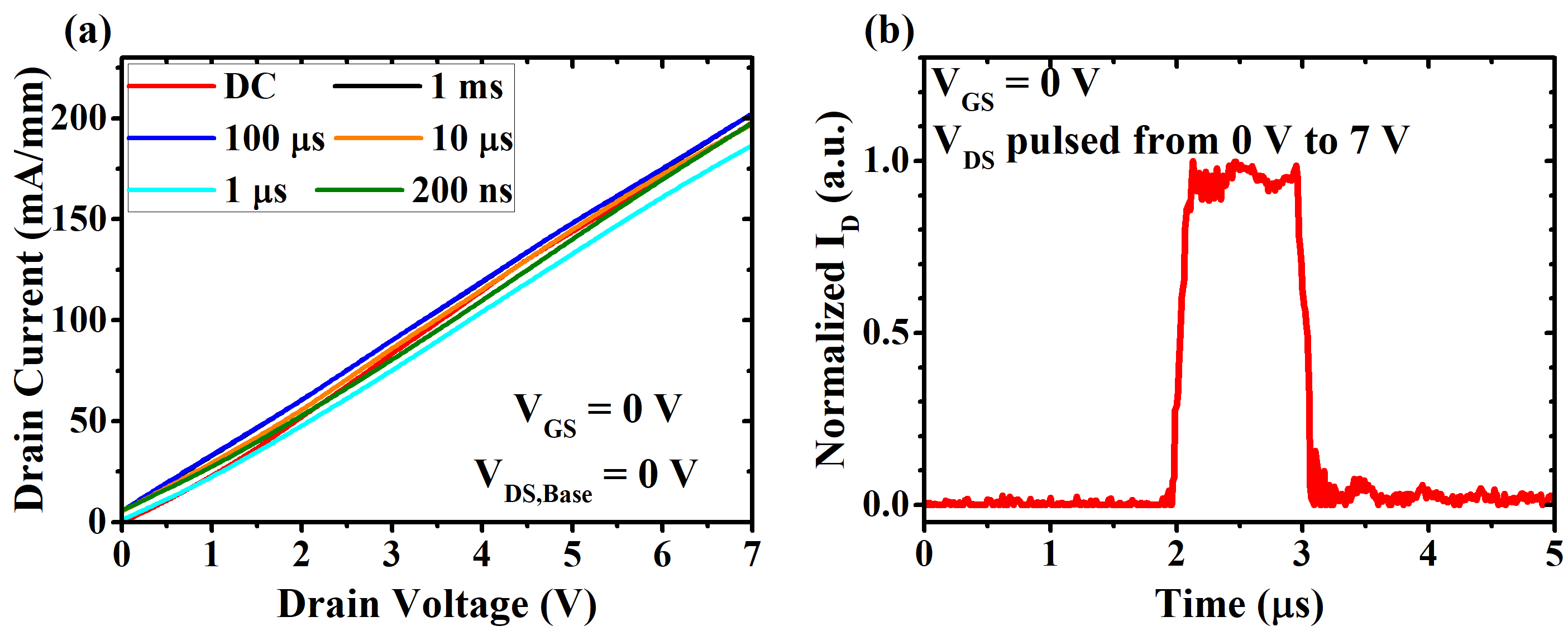}}
\caption{(a) Drain turn on I$_d$-V$_d$ curves for various pulse widths at V$_{gs}$ = 0 showing negligible current dispersion compared to Fig. \ref{id_vd_pulsed_tile}(b). (b)Drain turn on pulse with pulse width of 1~$\mu s$ showing no trapping effects which reduces the plausibility of bulk traps causing the current dispersion.}
\label{Dpulse_IdVd_1us_pulse_tile}
\end{figure}

\begin{figure}[b]
\centerline{\includegraphics[scale=0.175]{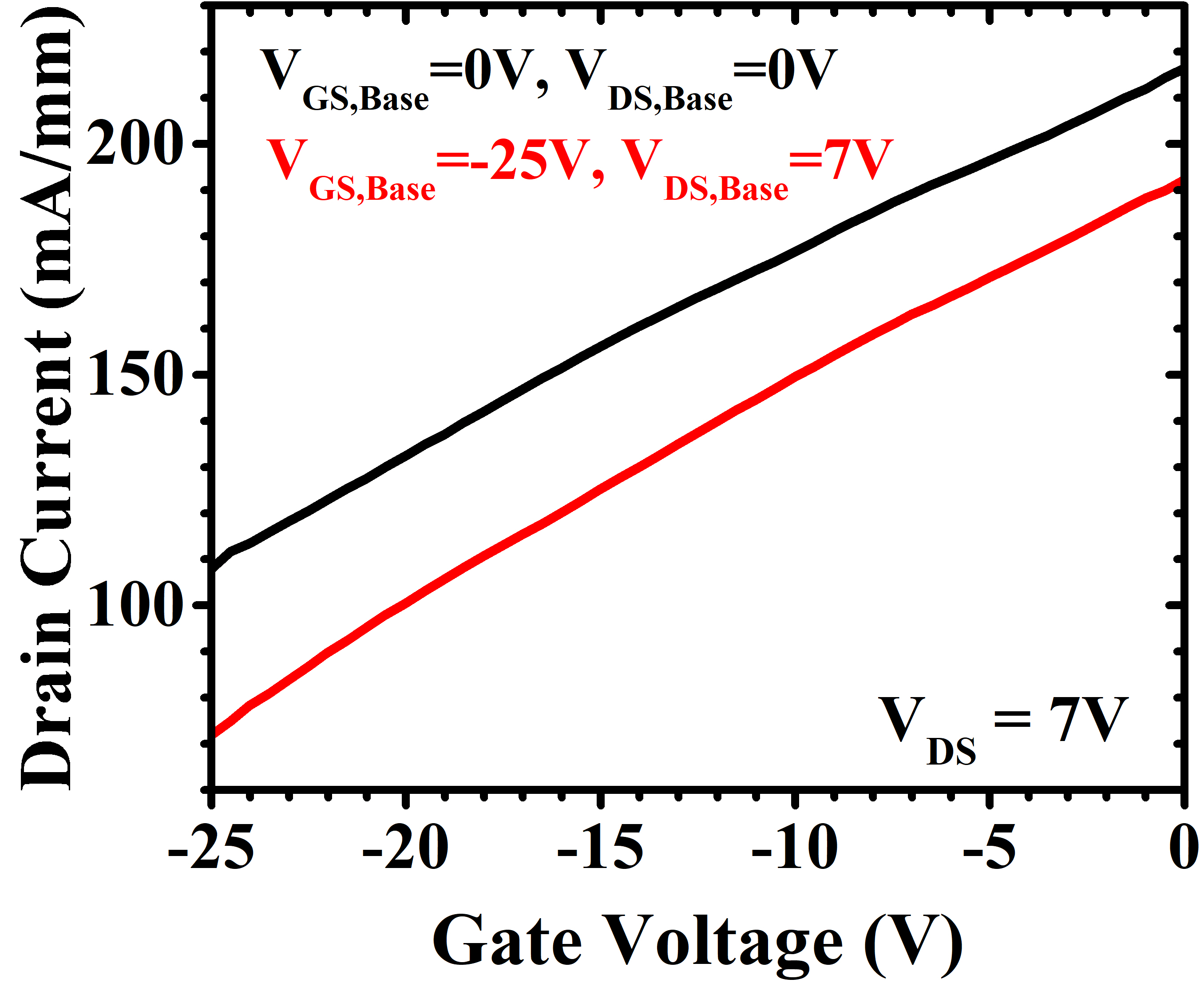}}
\caption{Double pulsed transfer curves using a 50~$\mu s$ pulse width for both gate and drain voltage pulses using the quiescent bias conditions as indicated in the figure. The black curve represents minimum trapping and hence more drain current levels while the red curve has reduced current due to trapping.}
\label{DbP_IdVg}
\end{figure}

\textit{Drain Turn On}: We first discuss the drain pulse measurements, as the bulk traps are expected to respond to drain pulse \cite{binari02,binari2002trapping}. Fig. \ref{Dpulse_IdVd_1us_pulse_tile} (a) shows device response to drain turn on pulse before passivation showing negligible dispersion in the I$_d$-V$_d$ plots for different pulse widths. Fig. \ref{Dpulse_IdVd_1us_pulse_tile} (b) shows the drain current pulse which does not show any delays in either transient edges. This rules out presence of any significant buffer traps. Fe diffusion from substrate has been identified as a trap in MBE grown Ga$_2$O$_3$ devices \cite{McGlone19}. However, the ozone MBE growth of the Ga$_2$O$_3$ does not show any appreciable Fe diffusion as seen in Fig. 3 in Ref.\cite{Wong15}. The same growth conditions are used in this study and the fabrication flow does not involve high temperature ($>$ 520$^{\circ}$C) processes. If both drain and gate are pulsed, we see change in the device characteristics as seen in Fig. \ref{DbP_IdVg}. This is due to the gate lag effect caused by interface traps which is discussed in the next section.

\begin{figure}[t]
\centerline{\includegraphics[width=\columnwidth]{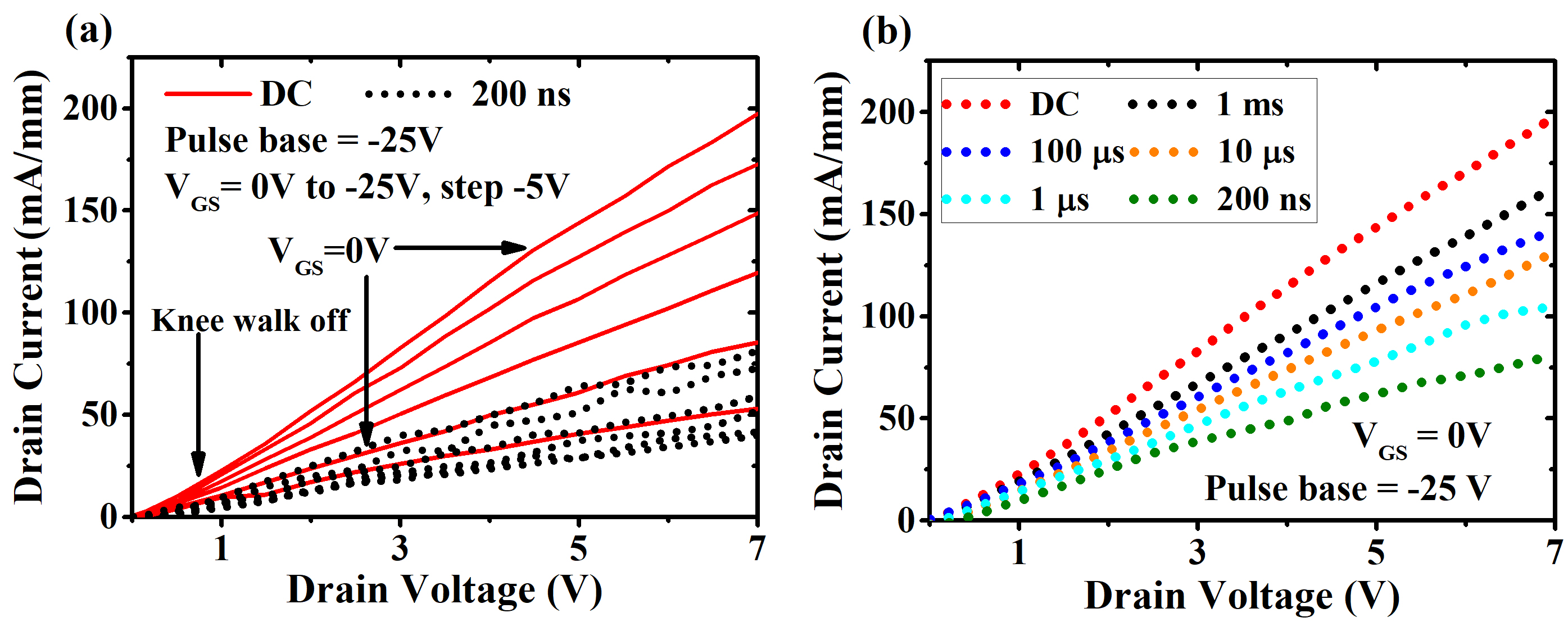}}
\caption{(a) Drain current profiles for V$_g$~=~0$V$ to V$_g$~=~-25$V$ for DC (red solid lines) and 200 $ns$ pulse width (black dotted lines), before passivisation. (b) Complete I$_d$-V$_d$ profile from DC to 200~$ns$ pulse width before passivation for V$_g$~=~-0$V$ shows clear dispersion in drain current with decreasing pulse width.}
\label{id_vd_pulsed_tile}
\end{figure}

\textit{Gate Turn On}: Figure \ref{id_vd_pulsed_tile}(a) shows DC and and gate pulsed I$_d$-V$_d$ characteristics of the unpassivated device, we can clearly see severe DC-RF dispersion and knee-walk off. The current collapse during gate turn on pulse could be related to both traps directly under the gate electrode and the traps in the gate to drain access region \cite{binari2002trapping}. Fig. \ref{id_vd_pulsed_tile}(b) shows the I$_d$-V$_d$ at a constant V$_{gs}$ as a function of pulse width. The current collapse increases with decreasing pulse widths. As discussed in the previous section buffer traps are ruled out. The dispersion is caused by interface traps under the gate and in the gate-drain access region.\par

\begin{figure}[b]
\centerline{\includegraphics[width=\columnwidth]{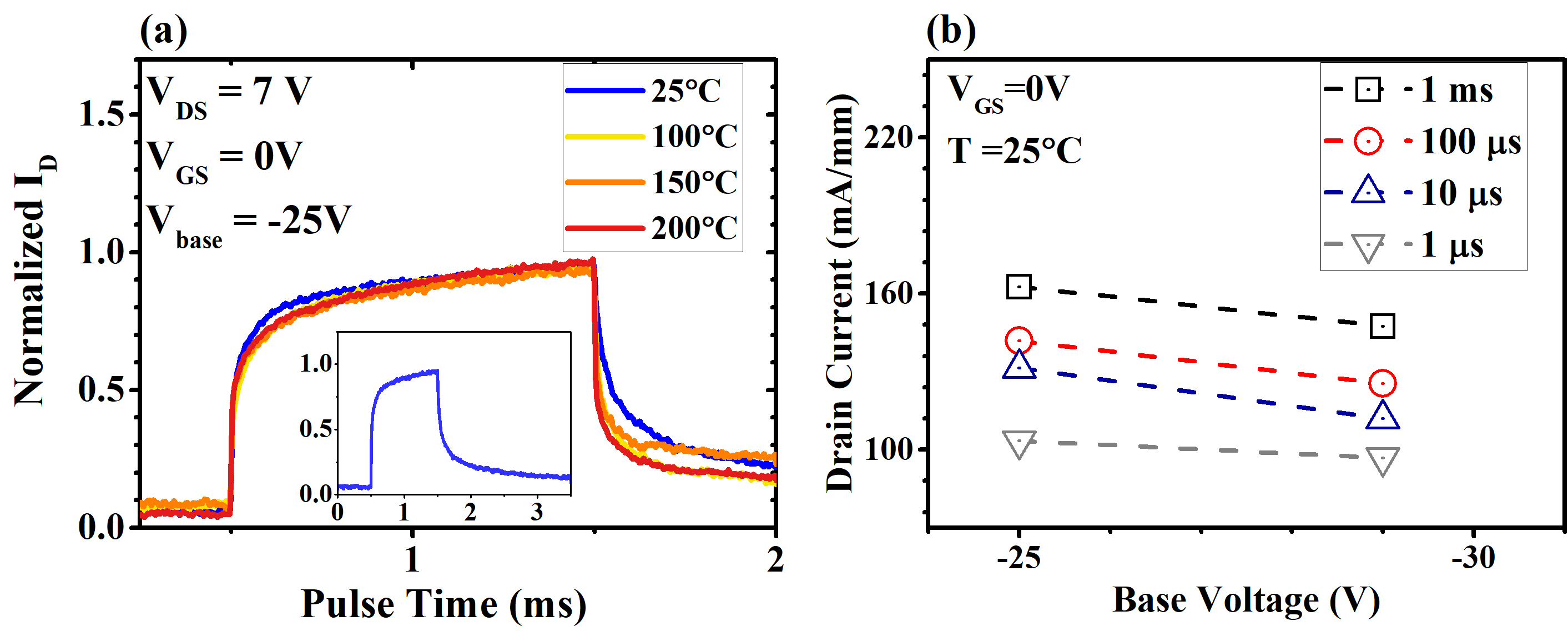}}
\caption{(a)Temperature dependent drain current pulse profile for a 1~$ms$ gate turn on pulse. Graph indicates how the traps response with temperature affects the drain current transient. Inset is room temperature plot for the same pulse width with extended turn off transient showing delayed effect due to capture time constants related to traps. (b) Change in maximum drain current recorded at V$_{gs}$~=~0$V$ depending on the base voltage of the pulse for varying pulse widths as indicated. A more negative base voltage assists more trapping and thus more current collapse.}
\label{1mspulse_basevtg_tile}
\end{figure}

\begin{figure}[t]
\centerline{\includegraphics[scale=0.175]{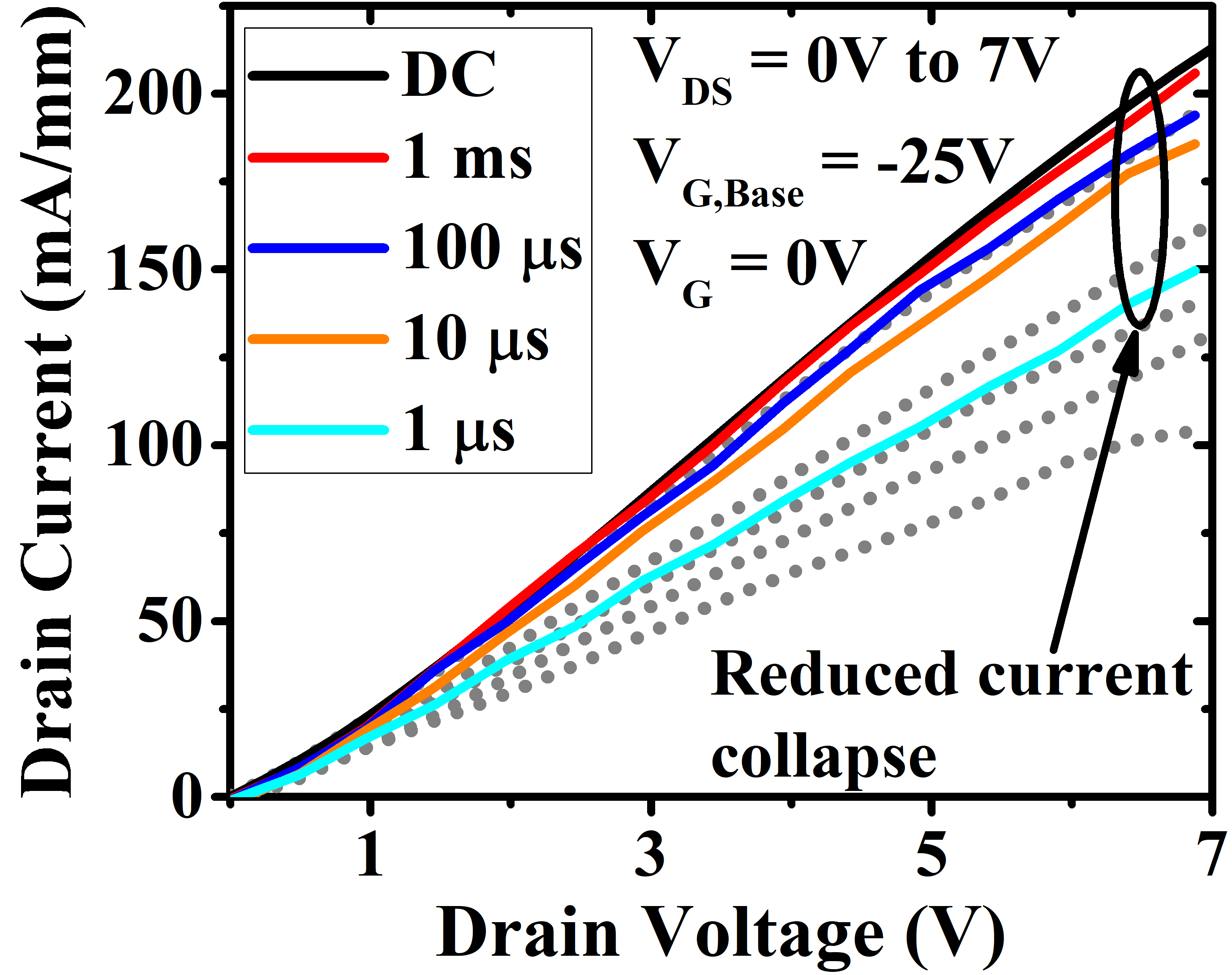}}
\caption{Complete I$_d$-V$_d$ profile from DC to 200~$ns$ pulse width before (dotted grey lines) and after (solid lines) passivation for V$_g$~=~-0$V$ shows significant improvement in drain current dispersion down to $10~\mu s$ pulse width.}
\label{IdVd_-25base_0Vg_all_pw_post_passv}
\end{figure}

\begin{figure}[b]
\centerline{\includegraphics[width=\columnwidth]{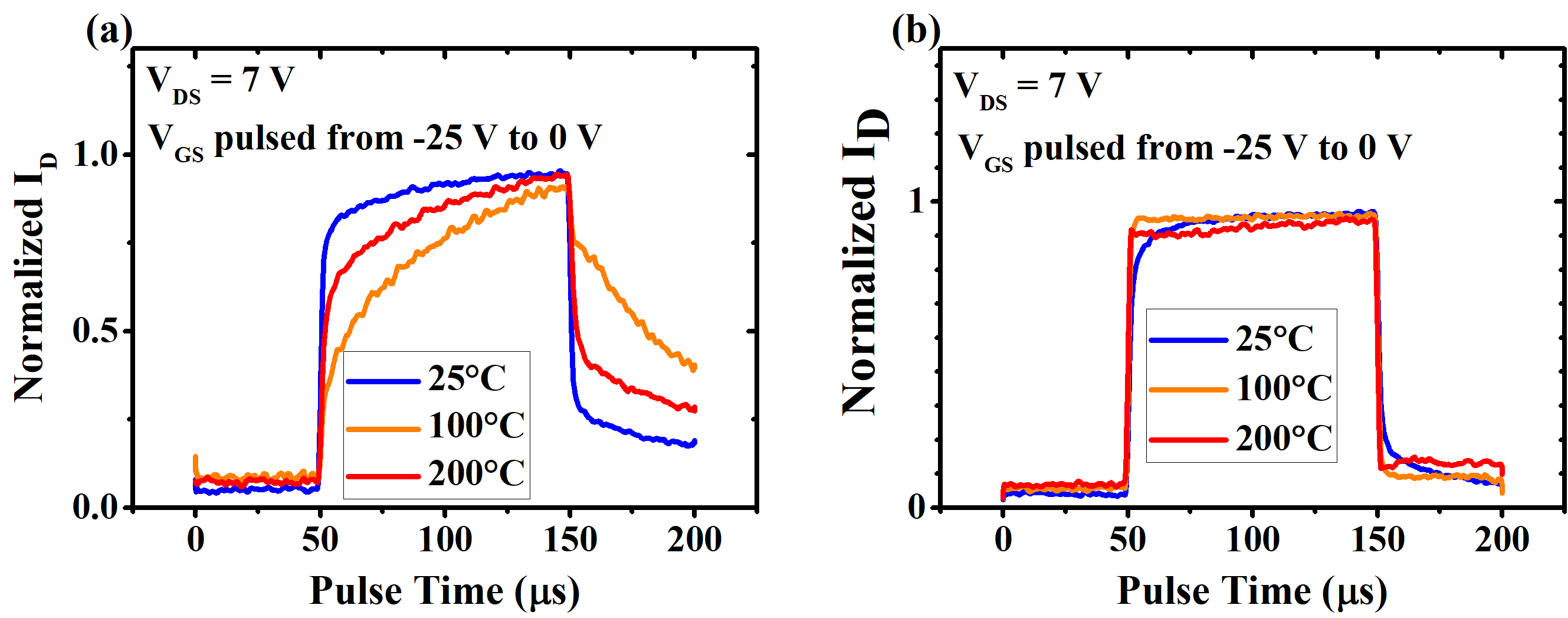}}
\caption{Temperature dependent drain current pulse profile for a 100~$\mu s$ gate turn on pulse before (a) and after (b) passivation.}
\label{100us_pulse_pre_post_passv_tile}
\end{figure}

Fig.\ref{1mspulse_basevtg_tile} (a) shows the temperature dependent time domain plot of drain current response to a 1~$ms$ gate turn on pulse. At all temperatures the drain current pulse shows a delayed asymptotic turn on transient which could be associated with emission time constants of electrons from the traps both in the gate-drain access region and the gate region. This pulse profile makes it self-explanatory as to why we see an increased  current dispersion between static and pulsed I$_d$-V$_d$ data with reduced pulse widths. A similar behaviour is seen during the turn off transient edge of I$_d$ pulse which could be attributed to capture time constants of electrons by the interface traps. Increasing the temperature, marginally affects the capture and emission time constants thus accounting for the slight difference in pulse shapes for different temperatures. Inset of Fig. \ref{1mspulse_basevtg_tile}(a) shows extended turn off transient for room temperature pulse to illustrate the slow capture process of electrons by traps after the gate voltage pulse returns to it's base value. As shown in Fig. \ref{1mspulse_basevtg_tile}(b), more negative the base voltage of the pulse, greater is the current collapse for the same pulse width. This is obvious since more negative base voltage would trigger more trapping of charges due to which drain current takes longer time to recover.\par 
\begin{figure}[t]
\centerline{\includegraphics[width=\columnwidth]{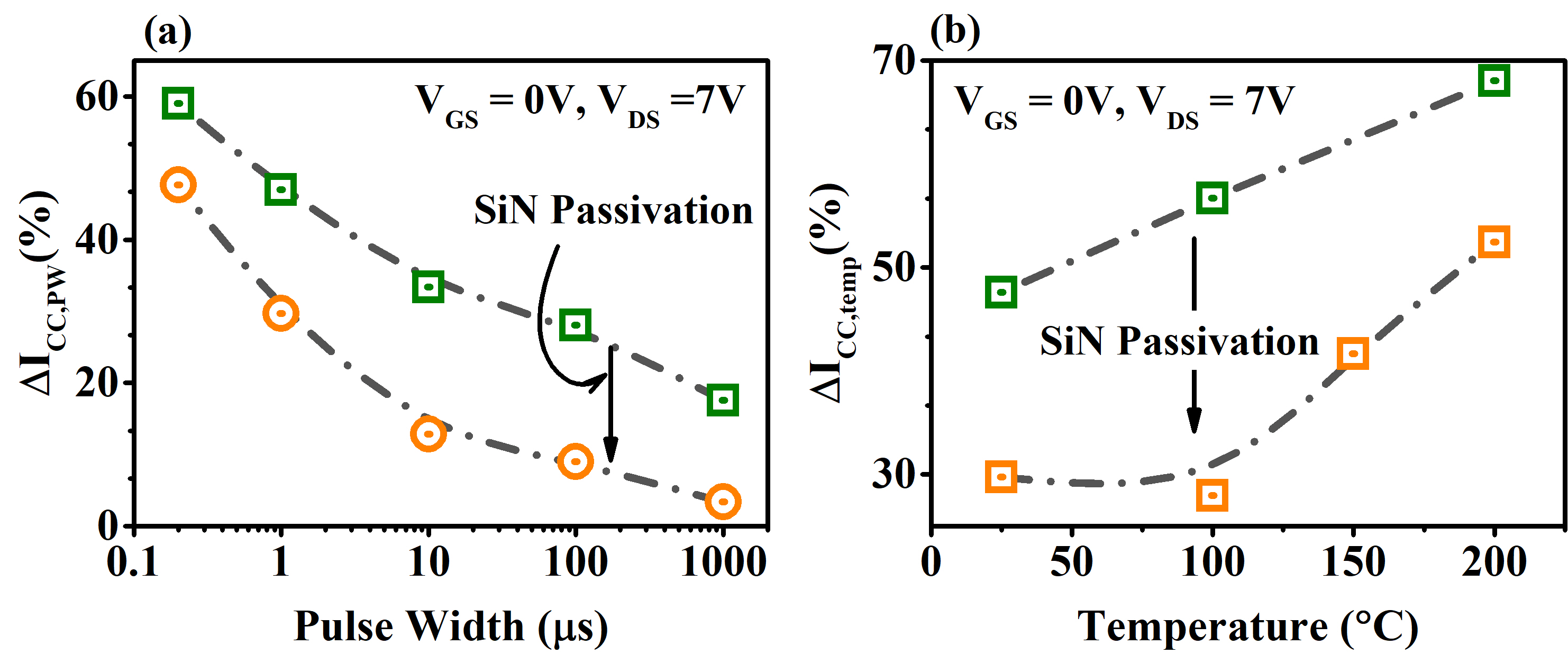}}
\caption{(a) Percentage drain current collapse with respect to DC measurement as a function of pulse widths before (green squares) and after (orange circles) passivation. (b) Percentage drain current collapse with respect to DC measurement as a function of temperature before (green squares) and after (orange circles) passivation.}
\label{Icc_temp_pw_tile}
\end{figure}

\subsection{SiN$_x$ passivation}\label{subsec:passivation}
According to the reports in previous studies on AlGaN/GaN devices \cite{binari02, Hashizume2003surface,Hyungtak2003effects, Green2000the}, SiN$_x$ passivation has helped create a near ideal semiconductor-dielectric interface by neutralizing the surface charge due to defects, dangling bonds and charged residuals. We explore the effect of SiN$_x$ passivated on RIE treated Ga$_2$O$_3$ devices.\par

Fig. \ref{IdVd_-25base_0Vg_all_pw_post_passv} compares pulsed I-V plots of the device post passivation with that of before passivation. We see a reduced DC-RF dispersion after passivation especially the low frequency dispersion is significantly reduced confirming that the role of traps in the gate-drain access region. While I$_d$-V$_d$ data corresponding to 1$\mu s$ and 200 $ns$ pulse widths still show dispersion, they both show considerable improvement in I$_d$ value over unpassivated device. The most likely reason for this dispersion is traps under the gate which are still in play. The temperature dependent drain current transients in Fig. \ref{100us_pulse_pre_post_passv_tile} clearly show the reduction of low frequency dispersion. Compared to the unpassivated drain current pulse, the passivated data is close to ideal square wave pulse shape except a very small delay at the end of the turn on and turn off transient which is responsible for the current dispersion seen in shorter pulse widths. High temperature pulse also shows a consistent improvement over its corresponding unpassivated pulse.\par 
To get a more quantitative idea of the effect of passivation on current collapse we have calculated percentage current collapse in relation to DC data as a function of temperature and pulse width using following equations:\\

\begin{equation}\label{current_collapse_pw}
    \Delta I_{cc,PW} = \frac{I_{d,DC} - I_{d,PW}}{I_{d,DC}} 
\end{equation}

\begin{equation}\label{current_collapse_temp}
    \Delta I_{cc,temp} = \frac{I_{d,DC} - I_{d,1\mu s}}{I_{d,DC}} 
\end{equation}

where $I_{d,DC}$ is static drain current value during DC measurement, $I_{d,PW}$ is drain current measured for one of the five pulse widths, and $I_{d,1\mu s}$ is drain current measured for 1$\mu s$ pulse. All the drain current values are measured at V$_{ds}$ = 7$V$ and V$_{gs}$ = 0$V$. As can be seen in Fig. \ref{Icc_temp_pw_tile} (a) current collapse is almost entirely recovered for larger pulse widths down to 10 $\mu s$ with collapse ratios below 10\% whereas the smaller pulse widths (1$\mu s$ and 200 $ns$) still show more than 30\% current collapse. The temperature dependent current collapse ratio for 1$\mu s$ pulse also shows average 20\% improvement over unpassivated data. The monotonous increase in current collapse with temperature is most likely attributed to higher rate of trapping at elevated temperatures. It is clear that the SiN$_x$ is effective in passivation of low frequency traps in the drain access regions. Further optimization of the passivation can completely eliminate the high frequency dispersion.\par

\subsection{Dynamics of traps}\label{subsec:trap_dyn}

Following the data and analysis in previous subsections we discuss the origin of the DC-RF dispersion. Since it is highly unlikely that as grown MBE substrate had a high concentration of surface traps, reactive ion etching step is responsible for the defect related interface traps observed in this study. A comparison with other published results shows that similar DC-RF dispersion was seen in \cite{Moser20} which used RIE for the channel layer. There has been report of RIE induced channel depletion in Ga$_2$O$_3$ MESFETs \cite{joishi2020deep}. While devices where the channel is not exposed to RIE \cite{Moser17, Singh18} show negligible DC-RF dispersion. Thus, the most likely cause of the dispersion in these devices are caused by RIE induced traps. During etching, $Ga$ vacancies are created which leave behind three dangling $O$ bonds that acts as a triple acceptor site according to McClusky \textit{et. al.} \cite{McCluskey20}. These traps form a virtual gate leading to depletion of electrons in drain access region which has slow turn on time when the gate is turned on.\par

The detrapping of electrons and hence the time response can occur by two possible methods: 1) thermionic emission from the trap, the rate for which is governed by the temperature of operation and/or 2) variable range hopping as described in \cite{Meneghesso13}. \par

\begin{figure}[t]
\centerline{\includegraphics[width=\columnwidth]{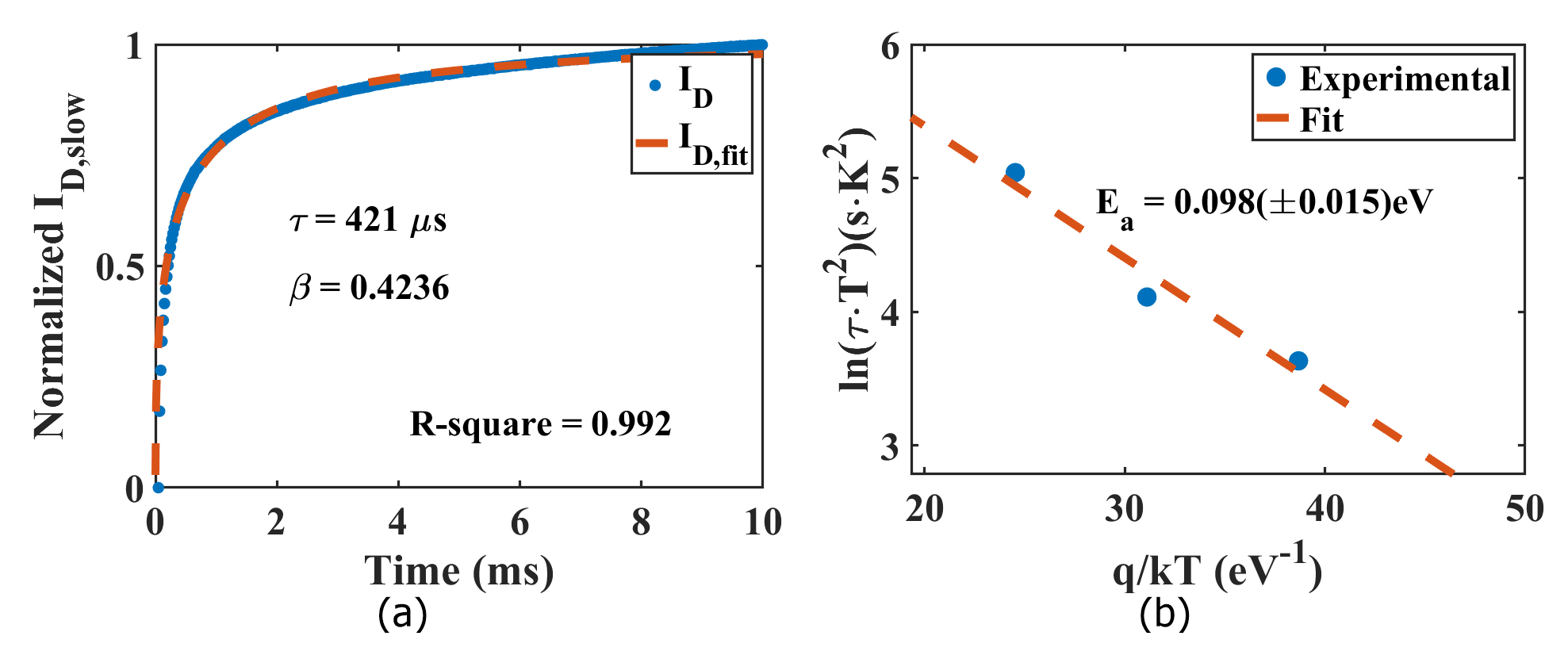}}
\caption{(a) A normalized $I_{D,slow}$ and a stretched exponential fitted curve gives a fitting parameter $\beta$ = 0.4236 and time constant $\tau$ = 421 $\mu s$. (b) Arrhenius plot of $q/kT$ vs $ln(\tau \cdot T^2)$. Slope of linear fitting line gives activation energy for the trap. }
\label{tau_fit_arrh_tile}
\end{figure}

Assuming a discrete or narrow band of traps in terms of energy level, the time constant of the traps can be determined by curve fitting a stretched exponential function in Eq. \ref{rev_expo_tau} to the drain current transient as shown in Fig. \ref{tau_fit_arrh_tile}~(a).
    \begin{equation}\label{rev_expo_tau}
        I_{d,slow} = 1 - exp \left(\frac{t}{\tau}\right)^\beta
    \end{equation}
    where $I_{d,slow}$ is the normalized slow transient, $\tau$ is the time constant and $\beta$ is the fitting parameter. A slow time constant of 421~$\mu s$  is obtained. From this time constant, the trap level is calculated \cite{Coffie03} using Eq. \ref{Ec-Et} .

    \begin{equation}\label{Ec-Et}
        E_c - E_t = kT\cdot\ln{\left(\frac{\tau~\sigma_n~\nu_n~g_1~N_c}{g_0}\right)}
    \end{equation}
where $E_c, E_t$ are conduction band edge and trap energy levels respectively, $\sigma_n$ is electron capture cross section of the trap, $\nu_n$ is the thermal velocity of electron, $g_0/g_1$ is degeneracy of the trap when it is occupied by 0/1 electron and $N_c$ is conduction band density of states. Assuming $\sigma_n = 10^{-15}~cm^2$, $\nu_n = 10^7~cm/s$, $N_c = 3.72\times10^{18}~cm^{-3}$ for this calculation. The calculated trap level is deep with an energy level of $0.41~eV$ below conduction band. However, a deep trap such as $0.41~eV$ would present decreasing time constant with increasing temperature according to Shockley-Read-Hall recombination theory \cite{shockley1952statistics}, which is not observed here (see Fig. \ref{rev_expo_tau} (b)).\par

An alternative mechanism suggested by Meneghesso \textit{et al} \cite{Meneghesso13} where the variable range hopping is the rate limiting step can explain the observed data. Following their approach, we use the Arrhenius equation fitting to $q/kT$ vs $\ln(\tau T^2)$ plot, which gives a lower activation energy to be $0.098\pm0.015~eV$ as shown in Fig. \ref{tau_fit_arrh_tile}~(b). The trap level estimated from Fig. \ref{tau_fit_arrh_tile}~(b) corresponds to a relatively shallow acceptor trap which along with a variable range hopping induced delay in capture and emission could be a more plausible explanation. Being a slow process VRH manifests itself as a slow turn on transient of the drain pulse with respect to time as seen in Fig. \ref{1mspulse_basevtg_tile}(a).  Further experiments such as DLTS are necessary to confirm the location and physical origin of the traps.

\subsection{High Frequency Characterization}\label{subsec:spar}

\begin{figure}[t]
\centerline{\includegraphics[scale = 0.24]{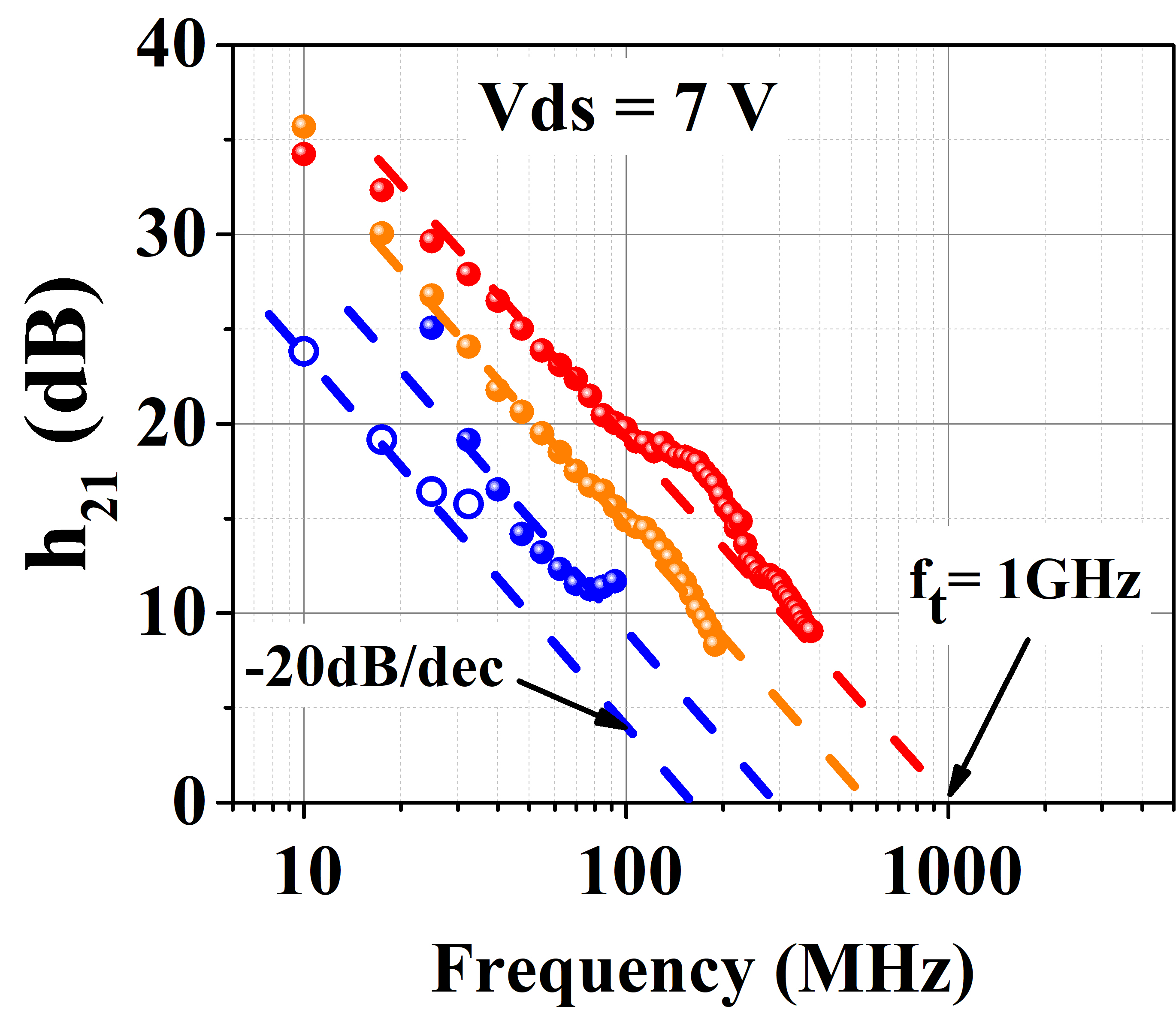}}
\caption{$h_{21}$ plots for various temperatures before passivation (hollow blue circles) and after passivation (spheres). Room temperature (blue spheres), 100$^{\circ}$C (orange spheres) and 200$^{\circ}$C (red spheres) indicate improvement in $f_t$ with increasing temperature.}
\label{rf_comparison_temp_dep}
\end{figure}

High frequency performance of the device was measured to primarily compare the data before and after passivation and to highlight the effects of surface traps on cutoff frequency ($f_t$). The device performance was degraded by high source and drain resistances and lower current modulation due to thicker channel in addition to trap induced increase in access resistance which is evident from the very low $f_t$ close to 100~$MHz$ before passivation in Fig. \ref{rf_comparison_temp_dep} (blue circles). Passivation improves the cutoff frequency by approximately two times which is due to reduced trapping in the access region that improves the transient characteristics of the device but could still be limited by the high source drain resistances and trapping directly under the gate unaffected by passivation. Temperature dependent $f_t$ measurement in Fig. \ref{rf_comparison_temp_dep} shows increase in cutoff frequency with temperature with a high $f_t$ close to 1~$GHz$ at 200$^{\circ}$C (red spheres). This could be attributed to both slightly decreased source and drain contact resistance and possibly high temperature assisted faster emission of trapped electrons under the gate.

\section{Conclusion}
\label{sec:conclusion}
We successfully fabricated a depletion mode Ga$_2$O$_3$ MOSFET with a 0.12~$\mu m$ gate length with a drain current of 210 $mA/mm$. Unpassivated devices show DC-RF dispersion with a peak current collapse ratio of 60\% at for 200 $ns$ gate turn on pulse. SiNx passivation show significant decrease in the current collapse. While the low frequency dispersion is eliminated the high frequency dispersion was present likely due to the traps under the gate. We also recorded a two fold improvement in cut-off frequency for room temperature measurements after passivation. A trap time constant of 421~$\mu s$ was obtained from stretched exponential curve fitting to the drain current pulse from the room temperature pulsed current-voltage measurements.\par
We have shown that RIE damage is likely responsible for the $Ga$ vacancies acting as acceptor like traps which induce a virtual gate in the access region between gate and drain which in turn is responsible for the current collapse observed in the transient response analysis of the device. The plasma damage also creates traps under the gate which modulate the threshold voltage of the device based on the temperature response of these traps. We also showed that the current dispersion can be mitigated, to some degree, by using a thick passivation layer of a silicon nitride. We investigated analytical response of the traps using stretched exponential curve fitting and provided a best fit explanation for the observed facts. Based on temperature dependent data, the variable range hopping based theory aligns much better with observed results than a pure Shockley-Read-Hall recombination theory.\par

\bibliographystyle{ieeetr}
\bibliography{references.bib}

\end{document}